\documentclass[reprint,superscriptaddress,amsmath,amssymb,aps,prx,floatfix,multicol]{revtex4-2}
\usepackage{bbm, bm, mathtools, braket, graphicx, hyperref, graphicx}
\usepackage{lipsum} 
\usepackage{blindtext}
\usepackage{xcolor}
\usepackage{graphicx}
\usepackage{dcolumn}
\usepackage{bm}
\usepackage{subfig}
\usepackage{soul}
\usepackage{float}
\usepackage{siunitx, nicefrac}
\usepackage{comment}
\usepackage{upgreek, multirow}
\usepackage{amsmath}
\usepackage{ragged2e}
\DeclareCaptionJustification{justified}{\justifying}
\newcommand{\ThreeJ}[6]{\left( \begin{matrix}  #1 & #2 & #3 \\ #4 & #5 & #6 \end{matrix} \right)}

\newcommand{\SixJ}[6]{\left\{ \begin{matrix} #1 & #2 & #3 \\ #4 & #5 & #6 \end{matrix} \right\}}

\begin{document}

\title{Narrowline Laser Cooling and Spectroscopy of Molecules via Stark States}% Force line breaks with \\

\author{Kameron Mehling}
% \email{}
\affiliation{
JILA, National Institute of Standards and Technology and the University of Colorado, Boulder, Colorado 80309-0440 \\ 
Department of Physics, University of Colorado, Boulder, Colorado 80309-0390, USA
}
\author{Justin J. Burau}

\author{Logan E. Hillberry}

\author{Mengjie Chen}

\author{Parul Aggarwal}
\affiliation{
JILA, National Institute of Standards and Technology and the University of Colorado, Boulder, Colorado 80309-0440 \\ 
Department of Physics, University of Colorado, Boulder, Colorado 80309-0390, USA
}
\author{Lan Cheng}
\affiliation{
Department of Chemistry, The Johns Hopkins University, Baltimore, Maryland 21218, USA
}
\author{Jun Ye}
\affiliation{
JILA, National Institute of Standards and Technology and the University of Colorado, Boulder, Colorado 80309-0440 \\ 
Department of Physics, University of Colorado, Boulder, Colorado 80309-0390, USA
}
\author{Simon Scheidegger}

\affiliation{
JILA, National Institute of Standards and Technology and the University of Colorado, Boulder, Colorado 80309-0440 \\ 
Department of Physics, University of Colorado, Boulder, Colorado 80309-0390, USA
}

\date{\today}

\begin{abstract}
The electronic energy level structure of yttrium monoxide (YO) provides a long-lived, low-lying $^{2}\Delta$ state ideal for high-precision molecular spectroscopy, narrowline laser cooling at the single photon-recoil limit, and studying dipolar physics with unprecedented interaction strength. High-resolution laser spectroscopy of ultracold laser-cooled YO molecules is used to study the Stark effect in the A$^{\prime}\,^{2}\Delta_{3/2}\,J=3/2$ state.  An immediate onset of the linear Stark effect is observed in the presence of weak applied electric fields due to the near degenerate $\Lambda$-doublet and the large electric dipole moment.  By applying a small electric field the Stark insensitive state is spectroscopically isolated and the absolute transition frequency to the X$\,^2\Sigma^+$ electronic ground state is determined with a fractional frequency uncertainty of $\num{9e-12}$. This electric field control is necessary to implement a quasi-closed photon cycling scheme that preserves parity. With this scheme the first narrowline laser cooling of a molecule is demonstrated, reducing the temperature of sub-Doppler cooled YO in two dimensions.
\end{abstract}
\maketitle
\section{Introduction}
Polar molecules are versatile quantum systems with a broad range of applications in the fields of precision spectroscopy~\cite{Gordon1954}, searches for fundamental symmetry violations~\cite{Hudson2002,ACME2018, Roussy2023}, quantum simulation of strongly interacting systems~\cite{Buchler2007,Goral2002,Lahaye2009,Pollet2010, christakis2023,miller2024}, quantum computation~\cite{DeMille2002,Yelin2006,Karra2016,Cornish2024, Zhang2022}, and cold chemistry~\cite{Ni2010,Bause2023}. The rich internal structure of polar molecules is foundational to these applications; exploitable features include narrow transitions over a broad spectral range, close lying states of opposite parity, and tunable intermolecular interactions. The complexity of the molecular structure makes it challenging to control the external and internal degrees of freedom. In pioneering works, experimental techniques have been developed to control the molecular motion using inhomogeneous electric and magnetic fields~\cite{Rabi1939,bennewitz1955fokussierung,ramsey1956molecular, bethlem1999decelerating,Meerakker2008,Reens2020,Jansen2020}, as well as optical pumping to address internal states~\cite{Drullinger1969,viteau2008optical,carr2009cold,Wu2020}. Today, direct laser cooling of molecules provides a route towards combined control on the single-quanta level.\newline
\indent The primary challenge of molecular laser cooling has been identifying sufficiently closed photon-cycling schemes. Early proposals suggested that vibrational closure is accessible in molecules with highly diagonal Franck-Condon factors~\cite{Rosa2004}. In addition, rotational closure is achievable when utilizing angular-momentum and parity selection rules between $\Sigma^+$ and $\Pi$ electronic states~\cite{stuhl2008magneto,Fitch2021}. Following this recipe, direct laser cooling of diatomic and polyatomic molecules has been achieved~\cite{zhelyazkova2014laser,Hummon2013,Hemmerling2016}, along with confinement in magneto-optical traps~\cite{Barry2014, Tarbutt2015, Anderegg2017,Collopy2018,Zeng2024,padilla-castillo2025}, sub-Doppler cooling~\cite{truppe2017molecules,Jarvis2018,ding2020sub,burau2023BDM,Hallas2024,Yu2024}, and optical dipole trapping with densities high enough to study two-body collisions~\cite{jorapur2023} at the single partial wave limit~\cite{Burau2024}. Whereas all these successes utilized strong $\Pi \leftarrow \Sigma^+$ and $\Sigma^+ \leftarrow \Sigma^+$ transitions, the use of other excited electronic states can further expand the internal and motional state control available to laser-cooled molecules.\newline
\indent Most laser cooled diatomic molecules possess a $^2\Delta$ state in the vicinity of the energetically lowest $^2\Pi$ state. If the $^2\Delta$ is the first excited electronic state (below the $^2\Pi$), the rotational closure of the $^2\Pi \leftarrow$ $^{2}\Sigma^+$ photon-cycling scheme is weakly broken. Spontaneous decay through the $^2\Delta$ state provides a two-photon decay pathway to opposite-parity ground states~\cite{Yeo2015}. Nevertheless, this electronic state configuration yields metastable $^2\Delta$ states with many unique features to utilize: $i)$ The long natural lifetime of the excited state allows for precise optical molecular spectroscopy; $ii)$ The narrow transition linewidth enables laser cooling to the photon recoil limit~\cite{collopy2015prospects, Truppe2019,Kobayashi2014}; $iii)$ The near-degenerate $\Lambda$-doublet and large molecular electric dipole moment $(\mu_{\text{e}}\approx 7\,\text{D})$ enable weak DC electric fields to polarize the molecules with orders of magnitude smaller field strengths than currently required for ground state molecules~\cite{Yi2007,matsuda2020resonant, Quemener2016, Schmidt2022}.
\newline
\indent Extending photon-cycling schemes to incorporate the metastable $^2\Delta$ state comes with two challenges related to rotational closure. First, the angular momentum of the $^2\Delta$ state is too large to establish rotationally-closed cycling as introduced by Stuhl et al.~\cite{stuhl2008magneto}. Radiative decay from the $^2\Delta$ excited state always branches to at least two $^2\Sigma^+$ rotational states. Second, the $\Lambda$-doublet splitting is so small that even stray electric fields partially polarize the $^2\Delta$ state. A polarized field-sensitive Stark state no longer possesses a definite parity, opening spontaneous decay pathways to four $^2\Sigma^+$ rotational states. State-of-the-art three-dimensional electric-field compensation is currently limited to approximately $\SI{0.5}{\milli\volt\per\cm}$~\cite{brandt2022measurement, clausen2021ionization}, insufficient to prevent partial polarization of the $^2\Delta$ state. In fact, it is much simpler to apply a small electric field to spectroscopically isolate a single field-insensitive Stark state which retains pure parity~\cite{Scheidegger2023}. \newline
\indent In this work we identify and implement photon-cycling schemes for the metastable $^2\Delta$ state within a laser-cooled molecule. To enhance rotational closure, we apply electric fields to control the $^2\Delta$ Stark states. These techniques are demonstrated with yttrium monoxide ($^{89}$Y$^{16}$O, shorthand YO) and are applicable to a broader class of metal-monoxides $M^{16}$O ($M = $ $^{45}$Sc, $^{89}$Y, $^{139}$La, $^{227}$Ac). The manuscript is structured as follows: In Sec.~\ref{Sec:Energy_level} we outline the low-lying electronic energy level structure of the $M$O molecules, focusing on the properties of the $^2\Delta$ state. Section~\ref{Sec:Experiment} introduces the experimental techniques used to prepare ultracold YO molecules. Section~\ref{Sec:Spectroscopy} presents the results on high-resolution optical spectroscopy of A$^{\prime}\,^2\Delta_{3/2}\leftarrow\,$X$\,^2\Sigma^+$ transitions. Section~\ref{Sec:Cooling} demonstrates the first narrowline laser cooling of a molecule (see Refs.~\cite{collopy2015prospects, Truppe2019,Kobayashi2014} for proposals).  Section~\ref{Sec: Prospect} discusses prospective experiments utilizing the longer-lived A$^{\prime}\,^2\Delta_{5/2}$ state. Finally, Sec.~\ref{Sec:Conclusion} presents conclusions drawn from this YO case study.\newline

\section{Energy level structure}\label{Sec:Energy_level}
\begin{figure}[h]
\includegraphics[scale=1]{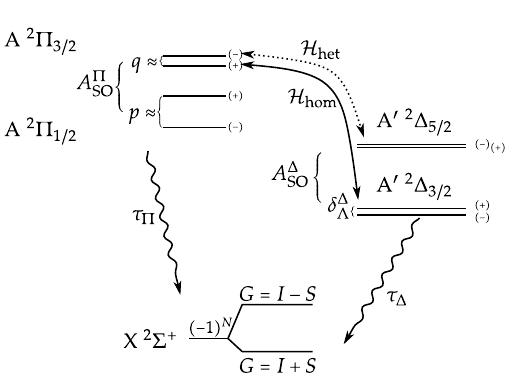}
\caption{Energy level diagram of the lowest three electronic states in the diatomic $M$O molecules. The large Fermi contact interaction dominates the fine structure in the X$\,^2\Sigma^+$ ground electronic state, whereas, the spin-orbit interaction ($A_{\mathrm{SO}}$) is dominant in the $^2\Pi$ and $^2\Delta$ excited states. The lifetime and the $\Lambda$-doublet splitting in the $^2\Delta$ state are primarily inherited from the $^2\Pi_{3/2}$ state through homogeneous $\mathcal{H}_{\text{hom}}$ and heterogeneous $\mathcal{H}_{\text{het}}$ interactions depicted by the double arrows. The parity of each state is depicted within the parentheses.}\label{Fig:1}
\end{figure}

\begin{figure}[h]
\includegraphics[scale=1]{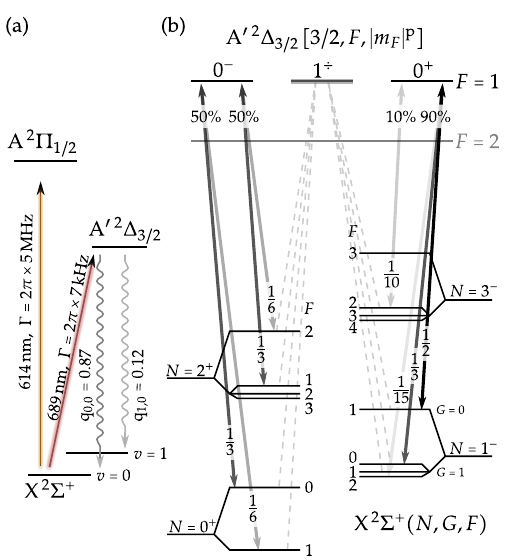}
\caption{(a) The relevant electronic states in YO with their transition wavelengths and radiative decay rates, along with the Franck-Condon factors q$_{v,v^{\prime}}$ between the first two vibrational states of X$\,^2\Sigma^{+}$ and the ground vibrational state of A$^{\prime}\,^2\Delta_{3/2}$. (b) The allowed decay channels for the A$^{\prime}\,^2\Delta_{3/2}[3/2,1,|m_F|^{\mathrm{p}}]\rightarrow$ ~X$\,^2\Sigma^{+}$  transitions. The branching ratios are provided and represented by the relative darkness of the double-arrow to each ground state. A weak electric field mixes the parity of $|m_F| =1$ states, leading to a breakdown of the parity selection rule for electric dipole transitions. Dashed lines represent the entirety of allowed decay pathways when the excited state has mixed parity.}\label{Fig:2}
\end{figure}

The common energy level structure of $M$O molecules arises from the strong ionic bond involving a single valence electron localized at the metal center. Figure\,\ref{Fig:1} schematically depicts the ordering of the lowest excited electronic states of the metal monoxides. They all have the same $\Sigma - \Delta - \Pi$ electronic state ordering, providing a metastable $^2\Delta$ state~\cite{Stringat1972,Childs1988,Yang2016,Bernard1980, zhang2020towards} with a natural lifetime much longer than the $^2\Pi$ state. The lowest electronic states have been well characterized and the spectroscopic constants for ScO, YO and LaO are available with sufficient accuracy for the broad transitions~\cite{Stringat1972, Childs1988,Yang2016,Bernard1979,Bernard1980,zhang2020towards,suenram1990a,Bernath2022,Bernath2023}. However, only a few studies of the long-lived $^2\Delta$ states are available~\cite{Chalek1976,Bernard1980,zhang2020towards} and important properties such as its $\Lambda$-doublet splitting remain unresolved. \newline
\indent The angular momentum coupling in the X$\,^2\Sigma^{+}$ ground state is dominated by the interaction between electron-spin and nuclear-spin, and best described using a Hund's case\,(b$_{\beta S}$) coupling scheme. The large Fermi contact interaction leads to two energetically well-separated manifolds with a total spin $\boldsymbol{G} = \boldsymbol{I}+\boldsymbol{S}$, where $\boldsymbol{S}$ is the total electron spin and $\boldsymbol{I}$ is the nuclear spin of the metal nucleus. The rotational angular momentum $\boldsymbol{N}$ couples with $\boldsymbol{G}$ to form the total angular momentum including spins $\boldsymbol{F} = \boldsymbol{N} + \boldsymbol{G}$. The notation $(N,G,F)$ will be used to specify the quantum numbers of the X$\,^2\Sigma^{+}$ state. This electronic ground state possesses definite parity, alternating between positive and negative parity according to $(-1)^N$ for a given rotational manifold.\newline
\indent The angular momentum coupling in the A$^{\prime}\,^2\Delta$ and A$\,^2\Pi$ excited electronic states is dominated by the spin-orbit interaction and best described by a Hund's case\,(a) coupling scheme. The term symbol $^{2S +1}|\Lambda|_{|\Omega|}$ describes the electronic state: $2S+1$ is the spin multiplicity, $\Lambda$ and $\Sigma$ are the projections of the electronic orbital angular momentum and electron spin onto the internuclear axis, respectively, and $\Omega = \Lambda + \Sigma$. The letters ${\Sigma, \Pi, \Delta,...}$ are used to represent $|\Lambda| = {0,1,2,...}$ in the term symbol. Since the projection of the angular momentum onto the internuclear axis can be positive or negative, there exist two quantum states ($\Lambda$ doublet) with identical magnitudes of $\Sigma$, $\Lambda$, and $\Omega$. The parity eigenfunctions for the Hund's case\,(a) basis are
\begin{equation}
    \left|\eta, \Lambda, \Sigma,\Omega, \pm\right> = \frac{1}{\sqrt{2}}\left(\left|\eta, \Lambda, \Sigma,\Omega\right> \pm \left|\eta, -\Lambda, -\Sigma,-\Omega\right>\right),
\end{equation}
where $\eta$ contains all other quantum numbers, such as the vibrational quantum number $v$, the total angular momentum quantum number $J$, the total angular momentum quantum number including spin $F$, and the projection quantum number $m_F$ required to define each quantum state~\cite{Brown_Carrington_2003,Watson2008}. \newline
 \newline
\indent The energy degeneracy of the $^2\Pi$ $\Lambda\,$doublets is lifted by interactions with nearby $^2\Sigma^+$ states. Since each rotational state in $^2\Sigma^+$ states has a fixed parity, a parity-dependent perturbation is induced through spin-orbit (SO) and electronic-rotational interactions~\cite{VanVleck1929,Mulliken1931a,lefebvrebrion04}. The perturbations result in a $\Lambda$-doublet splitting that strongly depends on the underlying coupling strength and the energy difference $\Delta E = |E_{\Pi} - E_{\Sigma}|$ between the electronic energy states. In the lowest rotational states of the A$\,^2\Pi$ electronic state, the $\Lambda$-doublet splitting can be estimated by the Mulliken parameter $q\approx 4B^2/\Delta E$ for $^2\Pi_{3/2}$~\cite{Mulliken1931a}, where $B$ is the rotational constant. Homogeneous SO interactions ($\Delta\Omega = 0$) are usually $A^{\Pi}_{\mathrm{SO}}/B \approx 10^3$ times stronger than heterogeneous electronic-rotational interactions ($\Delta\Omega = \pm 1$), where $A^{\Pi}_{\mathrm{SO}}$ is the spectroscopically resolved spin-orbit coupling constant of the $^2\Pi$ state. Thus, the $\Lambda$-doublet splitting for the $^2\Pi_{1/2}$ state can be estimated as $p \approx q A^{\Pi}_{\mathrm{SO}}/B$ ~\cite{Mulliken1931a}. The same mechanisms apply to the $\Lambda$-doublet splitting of $^2\Delta$ states via perturbation from the lifted degeneracy of the parity doublets within the nearby $^2\Pi$ states~\cite{Brown1987}.
\newline \indent The radiative lifetime $\tau_\Delta$ and the $\Lambda$-doublet splitting of the A$^{\prime}\,^2\Delta_{3/2}$ state are dominated by the admixed A$^2\Pi_{3/2}$ character resulting from the homogeneous interaction $\mathcal{H}_{\text{hom}}$. The class of metal monoxide molecules considered herein possess effective off-diagonal spin-orbit coupling matrix elements much smaller than the spectroscopic resolved spin-orbit coupling constant~\cite{zhang2020towards}, leading to lifetimes $\tau_{\Delta} \sim\SI{10}{\micro\second}$, about 10$^{3}$ larger than the $^2\Pi$ state lifetime $\tau_{\Pi}$. The admixture of the A$\,^2\Pi_{3/2}$ state determines not only the lifetime, but also the $\Lambda$-doubling splitting $\delta_{\Lambda}^{\Delta} \approx q \,\tau_{\Pi} /\tau_{\Delta} \sim q\cdot10^{-3}$ ($\delta_{\Lambda}^{\Delta}\approx \SI{6}{\kilo\hertz}$ for YO). The $^2\Delta_{5/2}$ state admixing with $^2\Pi_{3/2}$ is even further suppressed as it results from the heterogeneous interaction $\mathcal{H}_{\text{het}}$. Therefore, the radiative decay rate and the $\Lambda$-doublet splitting of the $^2\Delta_{5/2}$ state are expected to be at least $10^{3}$ smaller than in the $^2\Delta_{3/2}$ state. The dominant interactions for each electronic state and relative energy splittings are shown diagrammatically in Fig.~\ref{Fig:1}.\\
\indent The near degeneracy of the opposite-parity $\Lambda$-doublet states and the large molecular electric dipole moment (see Tab. \ref{Tab:Lifetimes} in Sec. \ref{Sec: Prospect}) makes the $^2\Delta$ states of $M$O molecules extraordinarily sensitive to external electric fields. The Stark interaction can be calculated precisely for the lowest $J$ states, where rotational distortion is negligible. The field dependent energy levels can be obtained by diagonalizing the effective Hamiltonian 
\begin{equation}
    \mathcal{H}_{\text{tot}} = \mathcal{H}_0 + \mathcal{H}_{\text{S}},
\end{equation}
where $\mathcal{H}_0$ contains the field free energy levels and $\mathcal{H}_{\text{S}}$ represents the non-relativistic Stark effect. The Stark effect matrix elements using parity symmetrized eigenfunctions (Eq. 8.429 from Ref.~\cite{Brown_Carrington_2003}) are calculated as
\begin{multline}\label{Eq:Stark full}
    \left<\Omega,J,I,F,m_F,{\pm}|-T^1_0(\boldsymbol{\mu}_{\text{e}})T^1_0(\boldsymbol{E})|\Omega,J',I,F',m_F,{\mp}\right> = \\
    -\mu_{\text{e}} E (-1)^{F - F' - m_F + I - \Omega}\Theta(F)\Theta(F')\Theta(J)\Theta(J')\\
    \ThreeJ{F}{1}{F'}{-m_F}{0}{m_F} \SixJ{J'}{F'}{I}{F}{J}{1}\ThreeJ{J}{1}{J'}{-\Omega}{0}{\Omega},
\end{multline}
where $\mu_{\text{e}}$ is the electric dipole moment of the $^2\Delta$ state, $E$ is the electric field strength, and $\Theta (x) = \sqrt{2x+1}$. Since the hyperfine splitting is much larger than $\delta_{\Lambda}^{\Delta}$ one can restrict the calculation in the weak-field limit to a single value of $\Omega$, $J$ and $F$. The weak-field limit of Eq.~\eqref{Eq:Stark full} simplifies to
\begin{equation}\label{Eq:Stark_approx}
\begin{aligned}
    \left<\Omega,J,I,F,m_F,{\pm}|T^1_0(\boldsymbol{\mu}_{\text{e}})T^1_0(\boldsymbol{E})|\Omega,J,I,F,m_F,{\mp}\right> =& \\
     -\mu_{\text{e}}\frac{J(J+1) + F(F+1) - I(I+1)}{2F(F+1)J(J+1)}|\Omega| E m_F \coloneq&\\ -\mu_{\text{eff}} E m_F .
\end{aligned}
\end{equation}
The Stark interaction induced by electric fields on the order of ${10^{-3}\,\unit{\volt\per\centi\meter}}$ is comparable to $\delta_{\Lambda}^{\Delta}$ and polarizes the molecule in the laboratory frame. The field dependent $|m_F|\neq 0$ states become aligned to the electric field and no longer possess definite parity. Hence, the parity selection rule for dipole transitions involving these states is no longer valid [see Fig.~\ref{Fig:2}~(b)].

\indent The spontaneous decay path A$^{\prime}\,^2\Delta\rightarrow\,$ X$^2\,\Sigma^+$ does not obey rotational closure, but one can limit the branching to two rotational manifolds of the ground electronic state. The decay pathways for YO are depicted in Fig.~\ref{Fig:2}\,(b), where we use [$J$,$F$,$|m_F|^{\text{p}}$] to specify all relevant quantum numbers of the $^2\Delta_{3/2}$ state and a parity label p$\,\,\in\{+,-,\div$\} for a positive, negative, or mixed parity state, respectively. The branching ratios from A$^{\prime}\,^2\Delta_{3/2}$[3/2, 1, $|m_F|^{\mathrm{p}}$] to the ground electronic state are calculated using the Hund's {case\,(b) to (a)} basis transformation provided in Ref.~\cite{Brown1976}. Spontaneous decays to even rotational states are equally distributed between $N=0$ and $N=2$. Conversely, the decays to odd rotational states possess $90\%$ branching to $N=1$ versus $10\%$ branching to $N=3$. Therefore spectroscopically addressing the positive parity A$^{\prime}\,^2\Delta_{3/2}\,[3/2,1,0^+]$ state is advantageous for establishing quasi-closed cycling. This is achieved by utilizing the Stark effect and demonstrated in Sec.~\ref{Sec:Spectroscopy}.

\section{Experimental Procedure and State Preparation}{\label{Sec:Experiment}}
All experiments in this study utilize an ultracold ensemble of $^{89}$Y$^{16}$O molecules prepared in a single hyperfine sublevel of the lowest two rotational states of X$\,^2\Sigma^+$. The experimental apparatus and sequence are described in Refs.~\cite{ding2020sub,burau2023BDM}, therefore we emphasize only key experimental details. A cold beam of YO with mean forward velocity of $\SI{160}{\meter\per\second}$ is produced by pulsed laser ablation of a ceramic Y$_2$O$_3$ target in a single-stage cryogenic helium buffer gas cell at $\SI{4}{\kelvin}$. The molecules are decelerated by chirped laser slowing utilizing the A\,$^{2}\Pi_{1/2}\,J=1/2\,\leftarrow \rm{X}\,^{2}\Sigma^{+}\,\textit{N}=1$ transitions. Sufficient cycling closure is achieved by optically repumping population branching to the first two excited vibrational modes ($v=1,2$) of the $\rm{X}\,^{2}\Sigma^+$ state. However, the $\Sigma-\Delta-\Pi$ electronic state configuration leads to weak violation of the rotational closure scheme. The low-lying $\mathrm{A}^{\prime}\,^{2}\Delta_{3/2}$ provides a two-photon decay pathway from $\mathrm{A}\,^2\Pi_{1/2}$ to the even parity $\rm{X}\,^{2}\Sigma^{+}$ $N=0,2$  electronic ground states. This leakage is resolved in the $v=0$ manifold by mixing the $N=1 \leftrightarrow N=0$ ground states with resonant microwave radiation and optically pumping the $N=2$ population into $N=0$. With this scheme, molecules are slowed to $<\SI{10}{\meter\per\second}$ and captured by a dual frequency magneto-optical trap (MOT), routinely loading $10^5$ molecules with a temperature of $\SI{2}{\milli\K}$. To further cool and compress the molecular cloud, a blue-detuned sub-Doppler MOT~\cite{burau2023BDM} is applied, followed by a gray-molasses-cooling pulse providing number densities of approximately $\SI{1e8}{\per\cubic\centi\meter}$ with a temperature of \SI{3}{\micro\kelvin}.
\newline
\indent Population transfer into a single quantum state $(N,G,F) = (1,1,0)$ is achieved by employing a near-resonant optical pumping sequence with two laser tones addressing the $\text{A}\,^2\Pi_{1/2}\leftarrow\text{X}\,^2\Sigma^+$ $(1, 1,\lbrace 1,2\rbrace)$ and $(1, 0, 1)$  transitions~\cite{Burau2024}. A 500-$\upmu$s long state preparation pulse transfers $\geq\,95\%$ of the population into the target state at the expense of marginal heating to $\SI{4}{\micro\kelvin}$. To mitigate further heating due to off-resonant scattering, the total intensity $I$ of each laser tone is set to $s = I/I_\mathrm{sat}\approx0.05$. The saturation intensity is $I_\mathrm{sat} = \pi hc\Gamma/3\lambda^{3}$ where $\Gamma$ and $\lambda$ are the natural linewidth and transition wavelength, respectively [Fig.~\ref{Fig:2}~(a)]. To initialize the molecules in an even-parity ground state, we transfer the $\text{X}\,^2\Sigma^+ (1, 1, 0)$ population to the $\text{X}\,^2\Sigma^+ (0, 1, 1)$ absolute ground state by applying an adiabatic Landau-Zener frequency-chirped microwave pulse. Following state preparation, all repump lasers and microwave tones are switched off unless otherwise stated.\\
\indent After state preparation, a tunable electric field $E_{\text{appl}}$ with a field strength $\SI{0.25}{\volt\per\cm}$ to $\SI{0.75}{\volt\per\cm}$ is applied along the $z$ direction opposing gravity. The weak applied electric field induces a Stark interaction between the near degenerate opposite-parity states in the A$^{\prime}\,^2\Delta$ manifold. The field strength is chosen such that the interaction is strong compared to the $\Lambda$-doublet splitting ($\delta_\Lambda^\Delta$), but weak with respect to the hyperfine splitting between the $F=1$ and $F=2$ A$^{\prime}\,^2\Delta_{3/2}\,J=3/2$ excited states. This electric field tunability enables precision Stark spectroscopy of the $\text{A}^{\prime}\,^2\Delta_{3/2}\,[J,F,|m_F|^{\mathrm{p}}]$ states using a \SI{689}{\nano\meter} external cavity diode laser. This narrow-linewidth laser is phase-locked to a low-noise erbium-doped fiber frequency comb, which is stabilized to an ultra stable monolithic cryogenic silicon cavity and referenced against the strontium clock transition~\cite{Oelker2019,Aeppli2024}. Circular-polarized light is coupled along the $x$-$y$ laser-beam axes to form a $\sigma^+\sigma^-$ optical molasse. Dark states are rapidly destabilized by the thermal velocity of the molecules~\cite{Papoff92, devlin2016three}. After a 4-ms long interrogation time we measure the remaining population of the initialized state (Sec.~\ref{Sec:Spectroscopy}) or the temperature of the molecular ensemble by time-of-flight expansion (Sec.~\ref{Sec:Cooling}) as a function of the applied laser frequency.

\section{Optical Stark Spectroscopy}\label{Sec:Spectroscopy}
\begin{figure*}[t!]
\includegraphics[scale=1]{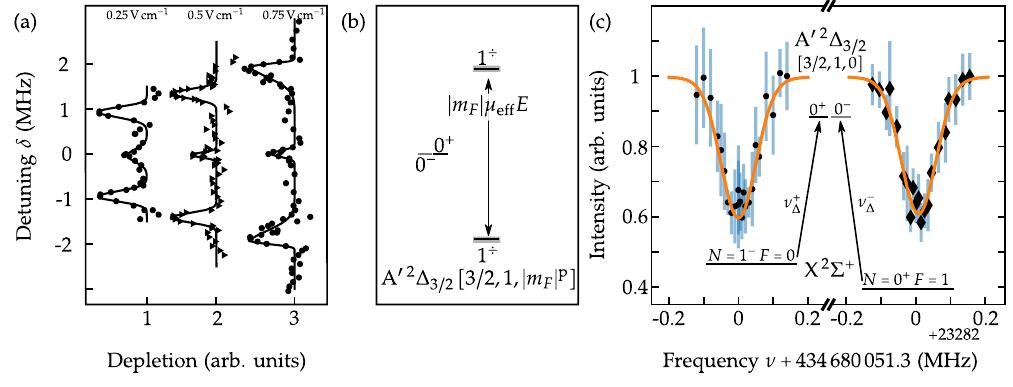}
\caption{Laser spectroscopy to the A$^{\prime}\,^{2}\Delta_{3/2}\,[3/2,1,|m_F|^{\mathrm{p}}]$ states in the presence of a weak applied electric field $E_{\mathrm{appl}}$. Panel (a) depicts the detuning-dependent depletion (solid markers) of the initial (1,1,0) ground-state population for three different values of $E_{\mathrm{appl}}$. Panel (b) shows the effect of the Stark interaction in the small field limit for the four field-sensitive and the two field-insensitive states. Panel (c) presents spectra from the $(1,1,0)$ (solid dots) and $(0,1,1)$ (solid diamonds) ground states to the two opposite pure-parity states of the A$^{\prime}\,^{2}\Delta_{3/2}\,[3/2,1,0]$ state. A Gaussian lineshape model (orange lines) is used to obtain each center frequency.}
\label{Fig:3}
\end{figure*}
\indent The long lifetime of the A$^{\prime}\,^2\Delta$ state provides narrow transitions to the electronic ground state while the small $\Lambda$-doublet splitting leads to a large electric field susceptibility. We investigated these properties by precision optical spectroscopy in the presence of a small applied electric field. The A$^{\prime}\,^{2}\Delta_{3/2} [3/2,1,|m_F|^{\mathrm{p}}]\leftarrow \text{X}\,^{2}\Sigma^{+}$ transitions are resolved by measuring depletion signatures of the ground-state population following a 4-ms long exposure to the near-resonant laser. Molecules remaining within the initial ground state are measured by a fluorescent readout scheme utilizing the strong A$^2\Pi\leftarrow$ X$^{2}\Sigma$ transition. Figure~\ref{Fig:3}~(a) presents the normalized depletion ratio of the $(1,1,0)$ population as a function of detuning $\delta$ from the field-free transition frequency. The depletion was enhanced by employing the two-tone $N=1$  optical pumping scheme via the broad transition to continuously repopulate the (1,1,0) ground state during the laser interrogation sequence. \newline
\indent Each spectrum consists of three depletion dips corresponding to transitions to different field-dependent Stark states. The data is fit by three skewed Gaussians accounting for Doppler and inhomogeneous Stark broadening. Figure~\ref{Fig:3}~(b) schematically illustrates the Stark interaction between the opposite-parity states of the $\Lambda$ doublet. The Stark interaction is proportional to the magnitude of the magnetic quantum number $|m_F|$ and the effective electric dipole moment $\mu_{\text{eff}}\approx 3.8\,\mathrm{D}$ (see Eq.~\eqref{Eq:Stark_approx} and Ref.~\cite{Smirnov2019}, respectively). The resonant frequencies of the Stark-sensitive $|m_F|=1$ states evolve linearly with the applied electric field strength [see Fig.~\ref{Fig:3}~(a)]. This linear field dependence demonstrates full polarization of the [3/2, 1,1] states for each applied field. Comparing the field-insensitive central transition frequency against the center-of-gravity of the field-dependent transitions places an approximate upper bound $\delta^{\Delta}_\Lambda\leq$ $\SI{30}{\kilo\hertz}$. Our experimentally constrained $\Lambda$-doublet splitting and the observed Stark shifts indicate the A$^{\prime}\,^{2}\Delta_{3/2}$ state is fully polarized by field strengths $E_{\text{pol}} \approx \SI{0.01}{\volt\per\cm}$.
\newline \indent The unperturbed central depletion dip within each spectrum corresponds to the $m_F=0$ transition frequency. This first-order Stark-insensitive state is used to define the field-free transition frequency and remains a pure parity eigenstate for all applied fields. Conversely, the field-dependent $|m_{F}|=1$ states no longer possess definite parity when polarized. The lack of definite parity explains the field-sensitive states' enhanced depletion within each spectrum, since the parity-mixed states possess additional decay pathways [dashed lines in Fig.~\ref{Fig:2}~(b)]. A weak electric field of approximately $\SI{0.5}{\volt\per\centi\meter}$ was applied throughout all following experiments. This applied field is sufficient to spectroscopically isolate the electric and magnetic field-insensitive $m_F=0$ states within the center of the Stark manifold.
\\
\indent Figure~\ref{Fig:3}~(c) presents ground state depletion measurements of both the [3/2,1,0$^+$] $\leftarrow$ (1,1,0) and the [3/2,1,0$^-$] $\leftarrow$ (0,1,1) transitions, with the corresponding frequencies denoted by $\nu^{+}_{\Delta}$ and $\nu^{-}_{\Delta}$, respectively. The ground-state populations are initialized as discussed above in Sec.~\ref{Sec:Experiment}. To mitigate line distortions, spectroscopy was taken at a low intensity (2\,$I_{\mathrm{sat}}$) while all other sources of radiation were switched off. The normalized depletion signals from $N=1$ (solid circles) and $N=0$ (solid diamonds) are presented with their corresponding standard deviations (blue lines) derived from 60 measurements. The centroid transition frequencies were obtained from a weighted least-squares Gaussian lineshape fit, accounting for the dominant inhomogeneous Doppler broadening. To recover the absolute transition frequencies $\nu^{+}_{\Delta}$ and $\;\nu^{-}_{\Delta}$, the measurements were corrected by subtracting the recoil and quadratic Stark shifts. Table~\ref{Tab:Corrections} summarizes the leading shifts and uncertainties considered for this absolute transition frequency determination.\\
\begin{table}[h]
	\caption{Frequency corrections $(\delta\nu)$ and uncertainties~$(\sigma)$ considered for determining the absolute transition frequencies $\nu^{+}_{\Delta}$ and $\;\nu^{-}_{\Delta}$.}\label{Tab:Corrections}\centering
	\begin{tabular}{lcc}
		\hline
		\hline
		\rule[-1.5mm]{0mm}{5mm}&$\delta\nu\,(\unit{\kilo\hertz})$ & $\sigma\,(\unit{\kilo\hertz})$ \\
		\hline
        Recoil shift & 3.99 & 0\\% needs to be subtracted from transition frequency
		Quadratic Stark shift &2.8 & 1.1 \\ %needs to be subtracted from transition frequency
        Doppler shift & 0 & 0.08\\
        \multirow{2}{*}{Zeeman shift} \hspace{2mm}$(1,1,0)$ & 0 & 0 \\
		\hspace{21.5mm}$(0,1,1)$ & 0 & 1.5 \\
        Pressure shift & 0 & $\sim(10^{-6})$ \\
        BBR ac-Stark shift & 0 & 0.1\\
        ac-Stark shift & 0 & $\sim(10^{-2})$ \\
		\hline
        Total & 6.8 & 1.9\\

	\end{tabular}
\end{table}

\indent The quadratic Stark shift is calculated precisely using Eq.~\eqref{Eq:Stark full}, the energy splitting to the nearest opposite-parity hyperfine state, and the observed first-order Stark shift of the field-sensitive states. The  A$^{\prime}\,^{2}\Delta_{3/2}\,J=3/2$ $F=2$ hyperfine state is measured to lie \num{105.3}(3)\unit{\mega\hertz} beneath the addressed $F=1$ state. The uncertainty of the quadratic Stark shift is estimated from the inhomogeneous line broadening of the field-dependent states. \\
\indent The first-order Doppler shift averages to zero by using a retroreflected, counter-propagating beam path. The associated uncertainty is derived assuming a maximal misalignment of \SI{1}{\milli\radian} between the counter-propagating beams and the velocity of the molecular cloud after \SI{6}{\milli\second} of free fall.\\ 
\indent The A$^{\prime}\,^{2}\Delta_{3/2}[3/2,1,0^{+}]\leftarrow$X$^2\Sigma^+ (1,1,0)$ transition has no first order differential Zeeman shift and the earth magnetic field is actively compensated in all three dimensions to less than 5~mG. The transition involving the X$^2\Sigma^+ (0,1,1)$ state is sensitive to a residual magnetic field and the Zeeman shift is estimated for a maximal population imbalance between the $m_F = \pm 1$ states of 10$\%$.\\
\indent The pressure shift, differential ac-Stark shift induced by black body radiation (BBR-shift), and ac-Stark shift induced by the spectroscopy laser are much smaller than the statistical uncertainty. The dominant contribution of the BBR-shift in the A$^{\prime}\,^{2}\Delta_{3/2}$ arises from the A\,$^2\Pi_{3/2}$ state and is estimated to be $\sim\SI{10}{\hertz}$ (following the procedure of Ref.~\cite{Farley1981}). The uncertainty for the BBR-shift is quoted to be 10 times this leading contribution, which is still well below our current experimental resolution. Close-lying hyperfine states are the leading contribution to ac-Stark shifts, which is calculated to be in the order of $\SI{10}{\hertz}$. The molecular density is too small for the van-der-Waals interaction to perturb the observed transition frequency, even under the assumption of partial polarization.\\
\indent Accounting for all shifts considered above, the corrected transition frequencies of the A$^{\prime}\,^{2}\Delta_{3/2} [3/2,1,0]\leftarrow \text{X}^{2}\Sigma^{+}$ transitions are:
\begin{align*}
\nu_{\Delta}^{+} &= \num{434680051302}(3)_{\text{stat}}(1)_{\text{syst}}~\unit{\kilo\hertz}\\
\nu_{\Delta}^{-} &= \num{434703333707}(3)_{\text{stat}}(2)_{\text{syst}}~\unit{\kilo\hertz}.
\end{align*}
These frequencies, with a relative uncertainty of \num{9e-12}, comprise two of the most precise direct optical electronic transitions measured within a polar molecule. Additionally, the difference between the two obtained optical frequencies $\Delta\nu = \nu_{\Delta}^{-} -\nu_{\Delta}^{+} = \num{23282405}(6)\unit{\kilo\hertz}$ is used to further constrain the $^{2}\Delta_{3/2}\:\Lambda$-doublet splitting. Comparing $\Delta\nu$ against the well-known first rotational splitting of the X$\,^2\Sigma^+$ ground states $\nu_{(0,1,1)}^{(1,1,0)}= \num{23282405}(4)~\unit{\kilo\hertz}$ obtained by Fourier-transform microwave spectroscopy~\cite{suenram1990a}, reveals $\delta^\Delta_\Lambda = \num{0}(7)\,\unit{\kilo\hertz}$. This result is in good agreement with the approximated value $\delta^\Delta_\Lambda \approx q \,\tau_\Pi / \tau_\Delta = \SI{6}{\kilo\hertz}$ derived in Sec.~\ref{Sec:Energy_level}.\\
\indent We demonstrated that excitation to a single quantum state of the A$^{\prime}\,^2\Delta_{3/2}$ state with a well-defined parity is possible in the presence of a weak applied electric field. Spectroscopic addressability of a pure-parity excited state is crucial for laser cooling on the narrowline transition to constrain the decay pathways to a minimal set of ground states. Furthermore, we used transitions between magnetic and electric field-insensitive states to measure optical transitions with an absolute accuracy \num{9e-12} and constrained the $\Lambda$-doublet splitting to below $\SI{7}{\kilo\hertz}$ by optical spectroscopy. 

\section{NarrowLine Laser Cooling}\label{Sec:Cooling}
The pure parity  [3/2, 1, 0] excited states provide the optimal A$^{\prime}\,^{2}\Delta_{3/2}$ cycling center for narrowline laser cooling.  Although no strict rotational closure scheme exists for $^2\Delta \leftarrow$ $^{2}\Sigma^{+}$ transitions, the asymmetric branching from [3/2, 1, 0$^{+}$] provides enhanced rotational closure within the $N=1$ manifold [see Fig.~\ref{Fig:2}~(b)]. However, photon cycling on a single sublevel of the A$^{\prime}\,^{2}\Delta_{3/2}$ excited state constrains the X$\,^{2}\Sigma^{+}$ (1,$G, F$) hyperfine sublevels available for laser cooling. Notably, $\Delta F=0$ transitions between $m_F=0$ sublevels are dipole forbidden, as well as $m_F^{\prime}=0\leftarrow m_F=\pm2$ transitions. Therefore, each hyperfine state within the X$\,^{2}\Sigma^{+}$ $N=1$ manifold possess dark sublevels except for the (1,1,0) state. For all subsequent data, the molecular population is  initialized in the (1,1,0) ground state as explained in Sec.~\ref{Sec:Experiment}. \newline
\indent  Narrowline laser cooling is demonstrated by employing the A$^{\prime}\,^{2}\Delta$[3/2, 1, 0$^+$] $\leftarrow$ X$\,^{2}\Sigma^{+}$(1,1,0) transition. Cycling closure within the $N=1$ manifold is achieved by continuously repopulating the (1,1,0) state. Two additional laser tones optically repump the (1,0,1) and (1,1,$\lbrace1,2\rbrace)$ ground states via A$^{2}\Pi_{1/2}$. Higher rovibrational branching is optically pumped back to $N=1$. This A$^{2}\Pi_{1/2}$-repumping and A$^{\prime}\,^{2}\Delta_{3/2}$-cooling scheme [right-hand side of Fig.~\ref{Fig:4}~(a)] is applied for 4 ms while the molecular cloud is in free fall. The measured radial temperatures presented in Fig.~\ref{Fig:4}~(a) (solid black squares) are obtained by time-of-flight measurements as a function of the static detuning $\delta$ from the narrowline transition resonance. The cooling transition linewidth is power broadened with a saturation parameter $s=I/I_{\text{sat}}\approx100$ to ensure the laser addresses broad classes of the molecular velocity distribution.
\newline
\indent The radial temperature of the molecular ensemble rises to $\SI{10}{\micro\kelvin}$ due to off-resonant scattering from the $N=1$ broadline repumpers. As the narrowline laser is scanned through resonance a strong temperature dependence is observed; heating is damped when the laser is red detuned ($\delta<0$) and enhanced when blue detuned ($\delta>0$). This radial temperature evolution is modeled by the time-integrated solution of the thermal rate equation~\cite{Foot2005}
\begin{equation}
\frac{\textnormal{d}T}{\textnormal{d}t} = 4\left(\frac{3+D}{3}\right)\frac{E_{\text{r}}R_{\text{scatt}}}{k_{\rm B}} - \frac{2\alpha}{m} T; \quad T(0) = T_0, 
\label{Diffusion}
\end{equation}
\noindent where $T$ is the radial temperature, $T_0$ the radial temperature in the absence of the narrowline laser, {$E_{\text{r}}$ the recoil energy ($E_{\text{r}}/k_{\rm B}= \SI{192}{\nano\kelvin}$), $m$ the mass of $^{89}$Y$^{16}$O, $R_{\text{scatt}}$ the scattering rate, $\alpha$ the damping coefficient, and $t$ the evolution time. The scattering rate and damping coefficient incorporate the detuning dependence according to $\Delta$=  $2\pi \times \delta$,
\begin{equation}
     R_{\text{scatt}} = \frac{\Gamma}{2}\frac{s}{1 +s+\left(2\Delta/\Gamma\right)^{2}},
     \end{equation}
     and
     \begin{equation}
     \alpha= -\frac{8\hbar k^{2} s}{\Gamma}\frac{\Delta}{\left[1+s+\left(2\Delta/\Gamma\right)^2\right]^2}.
\end{equation}
$\Gamma$ = $2\pi \times\SI{6.9}{\kilo\hertz}$ is the natural linewidth of the narrow transition~\cite{zhang2020towards}, and $k$ =  $2\pi$/\SI{689}{\nano\meter} is the corresponding wavenumber. Equation~(\ref{Diffusion}) is derived for $D$-dimensional cooling and three-dimensional isotropic spontaneous emission~\cite{Lett1989}. The model fit parameters for the two-dimensional cooling ($D=2$) are the saturation parameter $s$, the initial temperature $T_0$, and the evolution time $t$. The least-squares optimized values for the data presented in Fig.~\ref{Fig:4}~(a) are $s=220(30),$ $T_0 =\SI{9.87(12)}{\micro\kelvin}$, and $t = \SI{47(3)}{\micro\second}$. The qualitative agreement of the detuning-dependent temperature evolution across the $\SI{300}{\kilo\hertz}$ detuning range demonstrates the effect of narrowline Doppler cooling. Nevertheless, the photon scattering required to continuously optically pump the molecules to the (1,1,0) ground state outpaces the cooling on the narrowline transition. Therefore, any cycling scheme that implements repumping based on a broad transition provides cycling closure at the cost of heating.

\begin{figure}[h]
\includegraphics[scale=1]{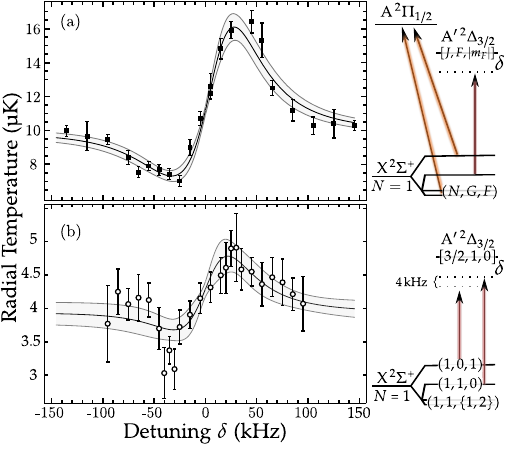}
\caption{Radial temperatures after applying a 4-ms long narrowline laser cooling sequence involving (a) repump lasers connecting the $^2\Pi_{1/2}$ state and (b) only narrowline transitions. A least-squares fit to the time-integrated solution to Eq.\ref{Diffusion} is provided (black line) with the $95\%$ confidence interval (gray shaded region) to model the detuning-dependent cooling for both scans. Each laser cooling scheme is depicted on the right-hand side.} \label{Fig:4}
\end{figure}

\indent To circumvent the repump-induced heating, we implement an alternative cooling scheme addressing only narrow transitions to the A$^{\prime}\:^{2}\Delta_{3/2}$ state. The [3/2, 1, 0$^+$] excited state predominately branches to two hyperfine levels within the $v=0$ manifold: 33.$\overline{33}\%$ to (1,1,0) and 25$\%$ to each of the (1,0,1) $|m_F|=1$ degenerate sublevels. With all molecules initialized in the (1,1,0) ground state, our two-tone narrowline laser cooling scheme [right-hand side of Fig.~\ref{Fig:4}(b)] provides closure for $83.\overline{33}\%$ of the rotational branching pathways. Accounting for higher vibrational loss, the number of photons scattered is limited to $\left< n_{\gamma} \right>$ = 3.6. Our all-narrowline laser cooling scheme is applied for 4 ms while the molecular cloud is in free fall, with each laser tone intensity set to $I\approx100 \,I_s$. The two tones are scanned in lockstep with a $\SI{4}{\kilo\hertz}$ two-photon detuning to prevent the formation of a $\Lambda$-enhanced dark state. In general, coherent dark states are rapidly destabilized at our thermal velocities by motion-induced non-adiabatic transitions~\cite{Papoff92} and do not limit the scattered photon number.
\newline \indent Figure~\ref{Fig:4}~(b) presents the measured radial temperatures (open black circles) as a function of $\delta$ from the [3/2, 1, 0$^+$] $\leftarrow$ (1,1,0) narrowline transition frequency. Even with a limited photon budget, the temperature evolution recovers a Doppler cooling signature. Having eliminated the off-resonant repump heating, the initial $\SI{4}{\micro\kelvin}$ radial temperature undergoes narrowline cooling (heating) as the laser is red (blue) detuned. The time-integrated solution to Eq.~\ref{Diffusion} is fit to the data according to a least-squares optimization (black line) with the accompanying 95$\%$ confidence interval represented by the gray-shaded region. 
\newline \indent The optimized model parameters are $s = 160(60)$, $T_0 = \SI{3.93(8)}{\micro\kelvin}$, and $t = \SI{13(3)}{\micro\second}$. The recovered saturation parameter and initial temperature agree with our independently measured values. To interpret the evolution time $t$, we derived the mean number of scattered photons $\left<n^{\prime}_\gamma\right>$. For our experimental configuration with two retro-reflected laser beams $\left<n^{\prime}_\gamma\right>=2\Gamma t = 1.1(3)$, which lies below our expectation of $\left<n_\gamma\right>=3.6$. The time-integrated solution to Eq.~\ref{Diffusion} models well the overall temperature evolution. However, the model deviates from the experimental data for detunings around $\delta = -\Gamma\sqrt{1+s}/4\pi$ where Doppler theory predicts the most efficient cooling \cite{loftus2004narrow}. The projected $\SI{250(160)}{\nano\kelvin}$ cooling underestimates the observed temperature reduction by more than a factor of two. The underestimated cooling deduced from the fit is self-consistent with the low number of scattered photons $\left<n^{\prime}_\gamma\right>$ and can be attributed to the relative simplicity of the model.
\newline \indent The measured mean temperature reduction of \SI{730(130)}{\nano\kelvin} across the five data points in the $\SI{-45}{\kilo\hertz}~$to$~\SI{-25}{\kilo\hertz}$ range provides clear experimental evidence for the first narrowline laser cooling of molecules. From the $\left< n_{\gamma} \right>~=~ 3.6$ photon number budget and the photon recoil energy, one can conclude that this narrowline cooling is very efficient. Further cooling is limited by the cycling closure and because the radial temperature approaches the Doppler temperature limit for the power-broadened transition linewidth. Both limitations are merely technical and can be overcome with modifications to our all-narrowline cooling scheme. \newline 
\indent The mean number of scattered photons can readily be enhanced by addressing all allowed decay pathways to the vibrational ground state ($v=0$), namely, the $10\%$ branching  to (3,1,2) and $6.\overline{66}\%$ branching to (1,1,2). The addition of two laser tones repumping these ground states via narrow transitions would double the number of photons scattered to $\left< n_{\gamma} \right>= 7.7$. Repumping the three $N=1$ hyperfine states within the $v=1$ manifold would enhance photon scattering to $\left< n_{\gamma} \right>= 45$. These prospective photon-cycling schemes require longer interrogation times, which can be achieved by loading the molecules into a state-insensitive magic-wavelength optical dipole trap~\cite{leung2020transition}. These future upgrades would enable cooling to the photon recoil limit or to the motional ground state within a tightly confining optical lattice or tweezer. Sideband cooling of trapped molecules~\cite{lu2024raman, bao2024raman} to their motional ground state will prove critical to improve current thermal-motion-limited rotational decoherence rates~\cite{burchesky2021rotational}.

\section{Prospects: $^{2}\Delta_{5/2}$}\label{Sec: Prospect} As described in Section \ref{Sec:Energy_level}, $M$O molecules possess a long-lived A$^{\prime}\,^2\Delta$ state with near-degenerate $\Lambda-$doublets. This excited state level structure enables weak electric fields to polarize the molecules and spectroscopically isolate individual Stark states (see Fig.\,\ref{Fig:3}). Although the experiments presented above primarily addressed the unperturbed pure-parity states, the A$^{\prime}\,^2\Delta$ field-sensitive states remain attractive
for future work in the fields of quantum simulation and precision measurements. To overcome the limitation of the A$^{\prime}\,^{2}\Delta_{3/2}$ lifetime, we propose to extend our experiments to the even longer-lived A$^{\prime}\,^{2}\Delta_{5/2}$ state in $M$O molecules. Relativistic exact two-component coupled-cluster singles and double (X2C-CCSD) 
and equation-of-motion CCSD (EOM-CCSD) calculations are used to obtain excited-state lifetimes, electric dipole moments, and nuclear Schiff moment sensitivity factors (see Supplemental Material for details \cite{Supplemental}\nocite{Matthews20a, CFOURfull,Dyall97,Ilias07, Liu09, Dyall01, Hess96a,Liu18, Visscher97, Stanton93a, Roos05, Roos05a, Kendall92, Liu18b, liu21}). \newline
\indent First, we consider the lifetimes of the A$^{\prime}\,^{2}\Delta$ electronic states for all $M$O molecules. These are calculated at the X2C-EOM-CCSD level~\cite{Asthana19} and are provided in Tab.\,\ref{Tab:Lifetimes}. Our calculations use the X$^2\Sigma^{+}$ state as the reference state and model the A$^{\prime}\,^{2}\Delta$ state by promoting the unpaired electron to the $\delta$ orbital.
\begin{table}[h!]
	\caption{Calculated lifetimes and molecular electric dipole moment for the first-excited metastable electronic states of the $M$O molecules.}\label{Tab:Lifetimes}\centering
	\begin{tabular}{lccc}
		\hline
		\hline
		\rule[-1.5mm]{0mm}{5mm}& $\tau_{\Delta_{3/2}} ({\unit{\milli\second}})$& $\tau_{\Delta_{5/2}} ({\unit{\milli\second}})$ &$\mu_{\mathrm{e}}({\text{Debye}})$\\
		\hline
        ScO &  0.795 & 954 & 8.1\\
		YO & 0.045 &  312 & 7.8\\
        LaO & 0.402 & 2,581 & 7.0\\
        AcO & 0.009 & 74 & 7.5\\
	\end{tabular}
\end{table}
The lifetimes of the A$^{\prime}\,^{2}\Delta_{3/2}$ states are determined by the A$^{\prime}\,^{2}\Delta_{3/2}\rightarrow$X$^2\Sigma^{+}$ transition. The transition dipole moments originate from the $n$s-$n$p hybridization in the X$^2\Sigma^{+}$ state and the $(n-1)$d-$n$p hybridization in the A$^{\prime}\,^{2}\Delta_{3/2}$ state due to the ligand field. The lifetimes of the A$^{\prime}\,^{2}\Delta_{5/2}$ states have contributions from two decay pathways, the A$^{\prime}\,^{2}\Delta_{5/2}\rightarrow$ X$^2\Sigma^{+}$ electric quadrupole transition and the A$^{\prime}\,^{2}\Delta_{5/2} \rightarrow $A$^{\prime}\,^{2}\Delta_{3/2}$ electric dipole transition. The general trend to shorter lifetimes for molecules with increasing atomic number $Z$ is attributed to both an increasing electric quadrupole moment of the A$^{\prime}\,^{2}\Delta_{5/2} \rightarrow$ X$^2\Sigma^{+}$ transition and the cubic scaling of the A$^{\prime}\,^{2}\Delta_{5/2} \rightarrow $A$^{\prime}\,^{2}\Delta_{3/2}$ decay rate with the spin-orbit coupling constant. One notable exception is that the long lifetime of LaO arises from the low transition frequency of A$^{\prime}\,^{2}\Delta_{5/2}\rightarrow$ X$^2\Sigma^{+}$, which leads to a reduction of the decay rate.
The calculated lifetimes are adequate to serve as an order-of-magnitude estimation, evidenced by the YO A$^{\prime}\,^{2}\Delta_{3/2}$ lifetime of \SI{45}{\micro\second} reasonably agreeing with the experimentally determined lifetime of 23(2)\,\unit{\micro\second}~\cite{zhang2020towards}. Our calculations suggest all A$^{\prime}\,^{2}\Delta_{3/2}$ states possess sufficiently narrow transitions to be used for resolved-sideband cooling (see Sec~\ref{Sec:Cooling}). Alternatively, the almost $10^4$ times longer A$^{\prime}\,^{2}\Delta_{5/2}$ lifetimes broaden the scope of experimental applications available to $M$O molecules. Although the field-insensitive A$^{\prime}\,^{2}\Delta_{5/2}\leftarrow$X$^2\Sigma^{+}$ ultranarrow transitions hold promise for a future optical molecular lattice clock, primary focus is paid to applications utilizing the electric field control of the long-lived A$^{\prime}\,^{2}\Delta_{5/2}$ field-sensitive states. \newline
\indent Quantum simulation with $M$O molecules in their A$^{\prime}\,^{2}\Delta_{5/2}$ field-sensitive states considerably relaxes the applied fields necessary to engineer long-range dipolar interactions. Traditionally, the strong intermolecular dipole-dipole interaction $V_{\text{dd}}$ is engineered by dressing adjacent rotational levels within the electronic ground state~\cite{Lahaye2009,Capogrosso2010,Pollet2010,Ciardi2025}. A lab-frame permanent dipole moment is induced as opposite parity rotational manifolds are coupled, either by strong dc-electric fields ($E_{\text{appl}} \sim {hcB}/\mu_{\rm e} \approx \SI{10}{\kilo\volt\per\cm}$)~\cite{matsuda2020resonant,Caroll2025} or near-resonant microwave radiation~\cite{miller2024,Yuan2025}. Conversely, the near-degenerate $\Lambda$-doublet and relatively large electric dipole moment of the A$^{\prime}\,^{2}\Delta_{5/2}$ state reduces the required DC electric field strength ($E_{\text{appl}} < {h\delta_{\Lambda}^{\Delta}}/\mu_{\rm e} \sim\SI{10}{\milli\volt\per\cm}$) by orders of magnitude. The strength of the long-range anisotropic dipole-dipole interaction can be tuned beyond $V_{\text{dd}}/h =\SI{10}{\kilo\hertz}$ in a \SI{1064}{\nano\meter} optical lattice, surpassing state-of-the-art molecular quantum simulation platforms~\cite{Blackmore2018,Holland2023,miller2024,Caroll2025,Yuan2025}. Furthermore, all A$^{\prime}\,^{2}\Delta_{5/2}$ state lifetimes referenced in Tab.\ref{Tab:Lifetimes} far exceed the interaction times necessary to observe intermolecular dipolar exchange~\cite{yan2013observation,ni2018dipolar}. Therefore, an alternative pathway to engineering strongly interacting quantum matter becomes realizable within the metastable A$^{\prime}\,^{2}\Delta_{5/2}$ excited state of $M$O molecules.   \newline
\indent Besides quantum simulation, polar molecules have also become widely adopted in experiments searching for violations of fundamental symmetries~\cite{Hudson2002,Aggarwal2018,DeMille2024,ACME2018,Roussy2023,Kozlov_1995,Safronova2018}.  When a charge-parity-violating interaction involves a time-reversal antisymmetric operator, e.g., in the search for an electron electric dipole moment (eEDM), the electronic contribution to the molecular sensitivity parameter comes entirely from the open-shell electrons. An enhanced sensitivity to an eEDM is obtained with a molecular state containing an $n\sigma$ molecular orbital~\cite{meyer2006candidate, Kozlov_1995}. Therefore, A$^{\prime}\,^{2}\Delta$ states are less sensitive than X\,${^{2}\Sigma^{+}}$ states, because the open-shell $n\delta$ electron contributes much less to the molecular sensitivity parameters than a core-penetrating $n\sigma$ electron. Although largely insensitive to the eEDM, the metastable A$^{\prime}\,^{2}\Delta$ states still prove suitable for other precision experiments searching for beyond-standard-model (BSM) physics.
\begin{table}[h]
	\caption{Molecular sensitivity factors for nuclear Schiff moment $W_{\text{NSM}}$ computed at the X2C-CCSD level for the X$\,^2\Sigma^{+}$ electronic ground state and the long-lived A$^{\prime}\,^2\Delta_{5/2}$ excited state of $M$O molecules  [in units of $\frac{e}{4\pi\epsilon_0a_0^4}\approx$ 44 $h\,\text{Hz}/(e\,\text{fm}^3)$].}\label{Tab:Schiff}\centering
	\begin{tabular}{lcc}
		\hline
		\hline
		\rule[-1.5mm]{0mm}{5mm}$M$O& \quad \text{$W_{\text{NSM}}$} & \quad \text{$W_{\text{NSM}}$} \\
         \rule[-1.5mm]{0mm}{5mm}& \quad $\left(\text{X}\,^2\Sigma^{+}\right)$ & \quad $\left(\text{A}^{\prime}\,^2\Delta_{5/2}\right)$ \\
		\hline
        ScO & \quad -309 & \quad -647\\
		YO & \quad -1,169 & \quad -2,444 \\
        LaO & \quad -1,794 & \quad -5,161\\
        AcO & \quad -23,110 & \quad -52,278\\
	\end{tabular}
\end{table}
\newline
\indent The A$^{\prime}\,^{2}\Delta$ states exhibit enhanced sensitivity for CP-violating interactions with molecular sensitivity parameters involving time-reversal symmetric operators. An important example, of rapidly growing interest, is the search for a nuclear Schiff moment (NSM) in the hadronic sector~\cite{Dzuba2002}. The NSM molecular sensitivity factor ($W_{\text{NSM}}$) has contributions from both closed-shell and open-shell electrons. Among the two types of chemical contributions to $W_{\text{NSM}}$~\cite{Chen2024}, the contribution due to the electron-drawing effects of the ligand comes from closed-shell electrons and is usually dominant. The back-polarization of the open-shell electron at the metal site provides an oppositely-signed contribution and reduces $W_{\text{NSM}}$. Importantly, this back-polarization contribution for $M$O molecules is much smaller
in the A$^{\prime}\,^{2}\Delta$ state than in the X$\,^2\Sigma^{+}$ state. The reduced back-polarization is attributed to the non-core-penetrating open-shell $n\delta$ electron. Enhanced molecular sensitivity factors are presented in Tab.~\ref{Tab:Schiff} where $W_{\text{NSM}}$ of the A$^{\prime}\,^{2}\Delta_{5/2}$ states are two-to-three times larger than the corresponding calculated values for the X$\,^2\Sigma^{+}$ ground states. This serves as the first proposal to use metastable A$^{\prime}\,^{2}\Delta_{5/2}$ excited states for precision searches of BSM physics. $^{227}$AcO is a particularly promising candidate for NSM searches because of the large octupole deformation of the $^{227}$Ac nucleus substantially increasing the measurable energy shift~\cite{Flambaum08,Verstraelen19,Flambaum25}. The resulting nuclear enhancement and large $W_{\text{NSM}}$ sensitivity factor of the $^{227}$AcO A$^{\prime}\,^{2}\Delta_{5/2}$ electronic state compares favorably against $^{227}$AcF ($W_{\text{NSM}}$ = -8,400 in the  X$\,^1\Sigma^{+}$ ground state), a molecule predicted to have record CP-violating NSM energy shifts~\cite{athanasakis2025laser}. 

\section{Conclusion}\label{Sec:Conclusion} 
All current laser cooled molecules rely on rotationally-closed cycling schemes exploiting highly diagonal and strong electronic transitions. While this recipe has enabled immense progress, the utility of other excited electronic states has largely remained unexplored. In particular, a subclass of molecules ($M$O) have a metastable A$^{\prime}\,^{2}\Delta$ as their first excited state, which possesses near-degenerate opposite-parity sublevels and long lifetimes. How these novel features provide additional quantum control for molecules is demonstrated with ultracold YO throughout this study.\newline
\indent We first characterize the narrow transition through optical spectroscopy of the  A$^{\prime}\,^{2}\Delta_{3/2}$ Stark states. Spectroscopy taken across a series of applied electric fields reveal full polarization in the laboratory frame at $E_{\text{pol}} \approx \SI{0.01}{\volt\per\cm}$. Transition frequencies to opposite-parity field-insensitive states were measured with a relative uncertainty of $\num{9e-12}$, constraining the A$^{\prime}\,^{2}\Delta_{3/2}$ $\Lambda$-doublet splitting to $\delta_{\Lambda}^{\Delta}\:\leq \SI{7}{\kilo\hertz}$. Pure parity excited states are isolated via the tunable Stark interaction and enable a quasi-closed photon-cycling scheme. \newline
\indent Utilizing our electric field control,  the first narrowline laser cooling of a molecule is achieved. The radial temperature of the molecular cloud was cooled by \SI{.73(13)}{\micro\kelvin} in free-space. Incorporating this novel cooling technique within an optical lattice is a promising pathway to initialize a large ensemble of molecules in their three-dimensional motional ground state for future quantum simulation platforms. \newline

\section{Acknowledgements}\label{Sec:Acknowledgment} 
 Funding support for this work is provided by AFOSR MURI, ARO MURI, NIST, NSF QLCI OMA-2016244, and NSF PHY-2317149. The computational work at Johns Hopkins University was supported by NSF PHY-2309253. We acknowledge A.~Aeppli, K.~Kim, D.~Lee, and B.~Lewis for maintaining the optical reference frequency used in this work. We thank T.~de Jongh and D.~J.~Nesbitt for careful reading of this manuscript.
\bibliography{main.bib}% Produces the bibliography via BibTeX.
\end{document}